\documentclass[11pt]{IEEEtran}

\usepackage{color}
\usepackage{graphicx}
\usepackage{amsmath}
\usepackage{wasysym}
\usepackage{multirow}
\usepackage{lineno}
\usepackage{booktabs}
\usepackage{array}
\usepackage{amssymb}
\usepackage{soul}
\usepackage[draft]{fixme}
\usepackage{longtable}
\usepackage{supertabular}
\usepackage{authblk}
\usepackage[colon]{natbib}
\usepackage{subfigure}
\usepackage[colon]{natbib}

\newcommand{\delOx}{$\delta{^{18}\mathrm{O}}$ }
\newcommand{\delD}{$\delta\mathrm{D}$ }

\begin{document}
\title{A continuous stream flash evaporator for the calibration 
of an IR cavity ring down spectrometer for isotopic analysis of water}

\author[1]{V. Gkinis}
\author[1]{T. J. Popp}
\author[1]{S. J. Johnsen}
\author[1]{T. Blunier}
\affil[1]{Centre for Ice and Climate, Niels Bohr Institute, University of Copenhagen, 
Juliane Maries Vej 30, DK-2100 Copenhagen, Denmark}

\maketitle

\section{abstract}
\noindent A new technique for high resolution simoultaneous 
isotopic analysis of \delOx and \delD in liquid water is presented.
A continuous stream flash evaporator has been designed
that is able to 
vaporise a stream of liquid water in a continuous mode and deliver a stable
and finely controlled water vapour sample to a commercially available Infra Red
Cavity Ring Down Spectrometer.

Injection of sub $\mu l$
amounts of the liquid water is achieved by 
pumping liquid water sample through a
fused silica capillary and instantaneously vaporizing it with a 100\% 
efficiency in a home made oven at a temperature of $170 ^{o}$C.
The system's simplicity, low power consumption and low dead volume together with 
the possibility for automated unattended operation, provides a solution
for the calibration of laser instruments performing isotopic analysis of water vapour.
Our work is mainly driven by the possibility to perform high resolution
on line water isotopic analysis on Continuous Flow Analysis systems typically
used to analyze the chemical composition of ice cores drilled in polar regions. 
In the following we describe the system's precision and stability, sensitivity
to varying levels of sample size and we assess the observed memory
effects. A test run with standard waters of different isotopic composition
is presented, demonstrating the ability to calibrate the spectrometer's
measurements on a VSMOW scale with a relatively simple and fast procedure.

%
%
%
\section{Introduction}

High precision stable isotope analysis of water
is typically performed offline via discrete sampling with traditional 
isotope ratio mass spectrometry (hereafter IRMS).  
While high precision and accuracy can routinely be achieved with IRMS 
systems, water isotope analysis remains an elaborate process, which is 
demanding in terms of sample preparation, power consumption, sample size, 
consumable standards,  and carrier gases.

In the most common IRMS techniques, water molecules 
are not measured as such, but are converted to a differentgas prior to measurement.
For \delOx analysis, the $\mathrm{CO_{2}}$ equilibration method 
\cite{Epstein1953}
has been widely used, whereas \delD analysis commonly involves the reduction of
water  to hydrogen gas over hot uranium
\cite{Bigeleisen1952, Vaughn1998}
or chromium
\cite{Gehre1996}.

However, combined use of these methods 
rules out simultaneous analysis of both water isotopologues 
on a given sample.  
More recently , in combination with 
continuous flow mass spectrometers, conversion of water to CO and H$_{2}$ is 
performed in a pyrolysis furnice
\cite{Begley1997, Gehre2004}
and allows simultaneous measurement, but still on a discrete sample.

Laser spectroscopy at the near and mid infrared regions
has been demonstrated 
as a potential alternative for water isotope analysis, presenting numerous 
advantages over IRMS
\cite{Kerstel1999, Iannone2009a}.
A major advantage of the technique 
is the ability to directly inject the sampled water vapour in the optical cavity 
of the spectrometer where both isotopic ratios $^{18}\textrm{O}/^{16}\textrm{O}$
and $^{2}\textrm{H}/^{1}\textrm{H}$
are simultaneuously measured.   Nowadays, commercial IR 
spectrometers  are available with a precision comparable to 
IRMS  systems
\cite{Lis2008, Brand2009}.
These units typically receive a continuous stream of water vapour sample and 
offer ease of use and portability.

The problem of the calibration of an IR spectrometer for isotopic analysis
of water vapour has been addressed in the past by other studies.
Wang et al
\cite{Wang2009}
calibrated an IR - spectrometer in an Off - Axis Integrated Cavity Output 
(OA - ICOS) configuration by monitoring the differences between a measured 
and a theoretically predicted Rayleigh distillation curve.
In order to obtain a saturated gas stream they make use of a dew point 
generator. 
This approach requires precise measurement of the total time of the Rayleigh
distillation process. The initial and final masses of the liquid water standard
used, also need to be measured. The time required for the process to complete
depends on the required precision and is of the order of 12-24 hours.

A different approach to the problem, is the generation  of water vapour with known
isotopic composition, by introducing liquid water standards in a dry gas stream,
 preferably at high temperatures. In this case, immediate 100\% 
evaporation is essential in order to 
avoid isotopic fractionation effects. Proper control of the injected amount of 
liquid water and the dry gas flow is essential for a stable water mixing ratio,
while it can in principle allow for tuning of the system to different humidity 
levels.
A dripper device, used by Lee et al
\cite{Lee2005}
introduced liquid water in a heated evaporation chamber filled with high purity 
nitrogen. The device operates in the range 800-30,000 ppm and delivers water vapour
sample to a direction absorption spectrometer operating in the mid IR region 
(6.66 $\mu m$).
Iannone et al
\cite{Iannone2009},
use a piezoelectric microdroplet generator 
\cite{Ulmke2001a}
in order to inject sub 
$\mu l$ amounts of water in a stream of dry gas where 100\% evaporation
takes place. The sample is then forwarded to a V-shaped high reflectivity 
cavity in an Optical Feedback - Cavity Enhanced Absorption Spectroscopy (OF-CEAS)
configuration
\cite{Morville2005, Kerstel2006}.
Aiming for in situ isotopic analysis of water vapour
in the Upper Troposphere - Lower Stratosphere region
where water mixing ratios are extremely low, the latter system is optimised
in the range between 12 and 3500 ppm.
Both evaporation systems use a relatively large volume, (order of mL)
static liquid water reservoir as a sample (or standard) source.

The motivation of this study lies in the area of ice core research with the goal to develop
a system that can perform on line \delD and \delOx isotopic analysis on a liquid sample 
stream originating from a continuously melting ice rod. 
Similar melting systems 
have been developed and used in lab or field environments for the analysis 
of chemical impurities 
or gases entrapped in the ice core
\cite{Fuhrer1993, Kaufmann2008, Schupbach2009}.
In this way, it is possible to perform measurements with high 
resolution when compared to equivalent measurements performed on discrete samples.
Portability, low power consumption and
cost have been parameters that were considered for this system.


\section{Experimental}

\subsection{NIR Cavity Ring Down Spectroscopy}
Water molecules present 
spectral absorption lines in the area of mid and near 
infra red due to ro-vibrational (fundamental and overtone) transitions.
At low sample pressures, the absorption lines are narrow enough to permit
distinction between different isotopologues.
These spectral features are unique and their relative intensities can  
be linked to relative isotopic abundances, hence providing the necessary
information to calculate isotopic ratios.
For a comprehensive description on the various optical techniques 
and analysis of the signals in IR spectroscopy as applied for isotopic analysis,
the reader may refer to 
\cite{Kerstel2004}.
In this study we use a commercial IR spectrometer purchased from
 Picarro Inc.
($L1102-i$)
\cite{Gupta2009}.
For high signal to noise ratio at a relatively low water concentration, the spectrometer utilizes a high
finesse cavity in a Cavity Ring Down (CRDS) configuration
\cite{Crosson2008}
. 
In a typical CRDS
experiment the laser light is coupled into the optical cavity and stays
in resonance until the intensity builds up to a maximum value. The light 
source is then turned off and the light intensity in the cavity decays 
exponentially as photons ``leak'' through the mirrors of finite reflectivity.
The result is that every photon completes thousands of roundtrips in the 
cavity, thus interacting with the injected absorbing sample through a path
length of the order of kilometers. The time constant of this decay, 
commonly refered to as ``ring down time'', depends
 only on the reflectivity of the mirrors $\mathcal{R}$, 
the length of the cavity $l_{c}$
and the absorption coefficient of the selected absorption feature 
$\alpha\left(\nu\right)$
\cite{Berden2000}.
The ring down time is described in Eq. \ref{Eq.3}.
Absorption spectra can thus be derived by the measurement of $\tau$ at 
different emmision wavenumbers $\nu$.

\begin{equation}
\label{Eq.3}
\tau \left( \nu \right) = 
\frac{l_{c}}{c \left[ \left( 1- \mathcal{R} \right) + 
\alpha \left( \nu \right) l_{c} \right]}
\end{equation}

In our system the IR spectrometer operates in a continuous flow-through
mode, maintaining a cavity pressure of 35 Torr at a gas
sample flow rate of $\approx30~mL/min$ at STP via two 
PID controlled proportional valves up and downstream of the optical cavity.
The temperature of the cavity is regulated at 80 $^{o}\mathrm{C}$.
The acquisition rate of the instrument is one
data point in $\approx$6 seconds.

\subsection{Sample preparation}
We implement a continuous liquid water stable isotope analysis by  
converting a stream of liquid water to water vapour and thereafter introducing the 
latter into the optical cavity of a NIR-CRDS spectrometer.
A sketch of the system is given in figure 1. Different liquid water  samples
(or standards) can be selected via a 6 port selection valve (V1). 
Transfer of the sample is done 
using a peristaltic pump that maintains a liquid flow rate
 of 0.1 $ml/min$ over the sample line. This flow rate can be 
adjusted according to the needs of the application or potential sample limitations.
High purity Perfluoroalkoxy (PFA) tubing with an ID of 0.5 mm is used in this
section.

A water level in the range of 17,000 - 22,000 ppmv
in the optical cavity results in  optimum spectrometer performance. 
With a nominal gas
flow rate of 30 $mL/min$ STP and a concentration
of 20,000 ppmv the required injection rate is $\approx 0.5 \mu L/min$ of liquid
water.
In order to introduce this- quantity of liquid water in the oven, 
we split off a fraction of the main sample line through a fused
silica capillary. The split takes place in a PEEK Tee split with \diameter 0.5 mm
bore (T1 in figure 1).

The small ID of the capillary tube acts as a restriction that imposes 
a back pressure at T1. In order to balance the capillary back pressure
we apply a restriction on the waste line downstream of the T1 Tee
by using tubing with an ID smaller than the \diameter 0.5 mm of the main sample line.
Assuming laminar flow, an estimate of the pressure build up along a tube is 
given by the Hagen-Poiseuille law as described in Eq. \ref{Poiseuille}.

\begin{equation}
\label{Poiseuille}
\Delta p = \frac{8\eta L Q}{\pi r}^{4}
\end{equation}
Where $\Delta p$ is the pressure drop along a tube
with length $L$ and inner diameter 2r and
$Q$ is the volumetric flow rate of the fluid with dynamic viscosity $\eta$.

Proper selection of length and inner diameter for the waste line and 
capillary tubes balances the two back pressures and offers a way to 
tune the flow rates through the capillary. At steady state  the flow
 through the capillary ($Q_{c}$) will be

\begin{equation}
\label{flow}
Q_{c} = Q_{w}\frac{L_{w}r_{c}^{4}}{L_{c}r^4_{w}} 
\approx Q_{m}\frac{L_{w}r_{c}^{4}}{L_{c}r^4_{w}}
\end{equation}
where $Q_{w}$ and $Q_{m}$ the flow of liquid sample at the waste and main
line respectively.
Due to the strong dependance of $Q_{c}$ on the inner diameter 
$\alpha$ of the tubes, the latter serve as a first order control on the 
flow through the capillary. 
Typical values for $\alpha_{c}$ and $\alpha_{w}$  are
20 $\mu m$  and 150 $\mu m$ respectively. Considering a fixed value for the
length of the capillary $L_{c}$ ($L_{cap} = 15 cm$), the linear 
dependance of $Q_{c}$ on $Q_{m}$ and $L_{w}$ allows for fine 
tuning of the flow by varying
the length of the tube at the waste line and the flow rate on the 
peristaltic pump.
A detailed view on the liquid sample handling and the tubing sizes involved 
is presented in figure 2.

The sample evaporation step is critical as 100\% 
immediate evaporation is essential in order to avoid isotopic 
fractionation effects. 
The evaporation oven consists of a a stainless steel Valco Tee-split 
 (Valco ZT1M) (T2 in figure 1),
attached on top of an aluminum block measuring $40 \times 40 \times 30$ mm.
The bore space in the Tee-split (\diameter 0.5 mm) serves as the evaporation chamber.
We maintain a temperature of 170 $^{o}$C  
by means of a PID controlled 200 W 
cartidge heater (Omega CSH-201200) fixed in the aluminum block. 
A temperature reading is obtained with a K type 
thermocouple. High temperature conductive paste is used in all 
of the connections of the evaporation chamber to ensure optimal 
heat distribution. 
Upon evaporation of the liquid sample in the oven, mixing 
with dry air takes place, forming the gas sample with the desired water vapour
levels.
Atmospheric air is dried through a \texttrademark Drierite canister
($\textrm{CaSO}_{4}$). Typical water levels for the dry gas are below 100 ppm 
water concentration.
For the transfer of dry gas to the evaporation chamber we use 
stainless steel 1/16" diameter tube.  
The mixture of dry gas and water vapor is transfered to the spectrometer 
at the nominal flow rate of 30 $cc/min$ 
via a 1/16" stainless steel tube that is heated to $\approx 90 ^{o}\textrm{C}$
in order to avoid recondensation of water vapour on the walls of the tube.

Overall our approach is towards minimizing dead volume in the system, thus 
reducing sample dispersion and limiting memory effects, while  at the same time we ensure 
efficient evaporation and negligible fractionation of the water sample.

%
%

\section{Results and Discussion}

\subsection{System Stability - Allan Variance}
We investigate the behavior of the spectrometer in combination with the sample
preparation system, regarding possible instrumental drift during long operation.
Injection of de-ionised water from a 5-liter bottle takes place for a period of 
$\approx$16 hours. The level of water vapour in the cavity during this period is
$19510 \pm 154 \textrm{ ppm } \left[1\sigma\right]$.
Results of the test are presented in figure 3. This test was performed 
with an acquisition rate of $\approx$1 data point every 6 sec. 
In a second stage we set the data on a fixed time step of 6 sec by means of  
linear interpolation.

We perform a ``clean-up'' of the raw data by removing outliers that deviate more 
than $\pm3\sigma$ from the mean value of the run.
For a total of 9551 points, 
we rejected 17 and 5 outliers for \delOx and \delD respectively. 
We observe that extreme outliers are likely to occur during
sudden and short term interruption of the water sample delivery. 
Small bubbles or other impurities
in the water stream are likely to cause such effects. Noisy and frequently 
interupted delivery of sample to the spectrometer can deteriorate the measuring
performance of the system.

Assuming that data points are normally distributed then a mean and variance
for this distribution are defined as:

\begin{align}
\label{Gaussian}
\bar\delta &= \frac{1}{N}\sum_{i=1}^{N} \delta _{i} \\
\sigma _\delta ^{2}&= \frac{1}{N-1}\sum _{i=1} ^{N} \left(\delta _{i} - 
\bar{\delta} \right) ^{2}
\end{align}

For sample size N the   standard error of the mean can be calculated as:
\begin{equation}
\label{variance_mean}
\sigma _{\bar{\delta}} ^{2} = 
\left(\frac{\partial \bar \delta}{\partial \delta} \: \sigma_{\delta} \right)^{2}
 = \frac{1}{N}\:\sigma_{\delta}^2
\end{equation}
Equation \ref{variance_mean} implies that in the theoretical case of a zero
drift system, one can progressively decrease the standard deviation of the mean
by increasing the integration time of the measurement. However, apparent
instrumental drifts are bound to limit the benefits of averaging a signal
over long integration times.

In order to assess the stability of the system we follow an approach similar to 
\cite{Werle1993}
and 
\cite{Czerwinski2009}
by calculating the Allan variance 
\cite{Allan1966}
for the time series presented in figure 3.
A time series of sample size $N$ can be devided in $m$ subsets of sample size
$k=\frac{N}{m}$. If the acquisition time per data point is $t_{i}$ then the 
integration time for every subset is $\tau_{m} = kt_{i}$.  
The Allan variance can then be defined as:
\begin{equation}
\label{Allan}
\sigma_{Allan}^{2} \left(\tau_m \right) = \frac{1}{2m}\sum_{j =1}^{m}
\left(\bar{\delta}_{j+1} - \bar{\delta}_{j} \right) ^{2}
\end{equation}
where $\bar{\delta}_{j+1}$ and $\bar{\delta}_{j}$ are the mean values 
of neighboring intervals $j$ and $j+1$.

In figure 4 we plot the calculated Allan variance  with different 
integration times for \delOx and \delD. For integration times up to about $\tau _{opt} = 5000$ sec
the Allan variance
decreases linearly for both isotopologues. This linear behavior suggests
a white noise signal and further averaging can lead to an 
improvement in the detection level of the system. However for integration times
longer than $\tau_{opt}$ averaging is not expected to improve the 
detection levels any further due to apparent instrumental drifts. 
Practically, for the purpose of calibration it is unlikeley that averaging times
longer than $\approx 600 $ sec will be implemented. As a result the optimum
performance
of the system at $\tau _{opt}$  is not fully exploited. 
However the value of $\tau_{opt}$ for both \delOx and \delD indicates a stable 
performance for both the spectrometer and the sample preparation line
with a precision comparable or better to mass spectrometry 
systems used for \delOx and \delD isotopic analysis. 
This behavior
is due to the optimized control of the spectrometer's cavity temperature and pressure
and the smooth and finely controlled evaporation and delivery of the water
vapour sample.

\subsection{Dependance on humidity levels}

We investigate the response of the system to different levels of injected water sample.
The pump rates of P1 are tuned in subsequent steps to 13 different humidity levels
spanning a range between 3.5 - 25 kppm. We acquire data for  $\approx$600 sec per
humidity level and select sections of these intervals according to their quality. 
The mean $\delta$ values of these sections are plotted as red and blue circles
for \delD and \delOx respectively at the bottom part of figure 5. The 
points are fitted with a 3$\mathrm{^{rd}}$ order polynomial plotted in black.
Error bars refer to $\pm1\sigma$ as calculated separately for every section. At
the top of figure 5 we present the water levels of a concatenated 
series of the different sections used. 

From the fitted curves in figure 5 we observe an overall linear response 
of \delOx to different water levels. This is however not the case for \delD , which shows
 a strong non-linearity at the low humidity area. For both isotopologues,
the uncertainty of the measurement shows a rapid increase below 5000 ppm.
In order to correct our measurements for possible fluctuations of the injected
water sample, we focus on the linear response area between 15 - 22 kppm. 
A linear fit on the data,  is presented in figure 6.
For each data point
the correction term $\Delta\delta_{hum}$ will be given by:

\begin{equation}
\label{Eq.4}
\Delta \delta_{hum} = \lambda \cdot (R_{20} - 1)
\end{equation}

Where $R_{20} = \frac{\left[\textrm{H}_{2}\textrm{O}\right]_{ppm}}{20000_{ppm}}$ 
and $\lambda$ is estimated from the linear regression:

\begin{equation}
\lambda_{18} = 1.94 \textrm{ }\permil\qquad
\lambda_{D} = 3.77\textrm{ }\permil\\
\end{equation}
So for a deviation of 1000 ppm
\begin{equation}
\Delta\delta ^{18}\textrm{O} = 0.097 \textrm{ }\permil \qquad
\Delta\delta \textrm{D} = 0.19 \textrm{ }\permil
\end{equation}

Brand et al \cite{Brand2009} have performed a similar calibration using  
water samples injected in a discrete mode 
with a commercially available sample 
preparation line that consists of an autosampler and a liquid vaporizer
\cite{Gupta2009}.
A comparison of the two data sets and the calibration lines shows a good agreement over
the whole range of humidities.

\subsection{Memory effects}
The continuous flow of liquid and gaseous
 sample in the transfer lines, the 
evaporation chamber and the optical cavity of the spectrometer
result in apparent dispersion effects. These effects impose a cross talk 
between samples commonly refered to as ``memory effects''.
In the case of isotopic analysis performed in a discrete mode the memory 
effect influencing the j$^{\textrm{th}}$ analysis of the run depends on the
isotopic value of the n previous analyses weighted by a set of memory
coefficients
So, if the expected value for the $\mathrm{j^{th}}$ analysis is 
$\delta _{j}^{r}$ and the measured equivalent is $\delta _{j}^{m}$, then the
memory effect $\mathrm{M_j}$ is described as:
\begin{equation}
\begin{aligned}
\label{memory_discrete1}
\delta _{j}^{r} =\delta _{j}^{m} + M_{j}\\
M_{j} = \sum _{k=1}^{n} \varphi _{k}(\delta _{j} - \delta _{j-k})
\end{aligned}
\end{equation}
Determination of the memory coefficients is used to characterize the 
experimental system and to correct 
the measured isotopic values for observed memory effects.

In the case of continuous measurements we follow a slightly different approach.
We generate an isotopic step by switching between two standard waters with 
different isotopic composition. This results in a smoothed sigmoid curve 
(figure 7) .
Ideally, in the case of zero
dispersion, a switch between two standards would be described by a scaled and 
shifted version of the the Heaviside unit step function as:
\begin{equation}
\label{Eq.6}
S\left(t\right) = 
\begin{cases}
C_{2} & t<0\\
C_{1} H \left(t \right) + C_{2} & t\geq 0
\end{cases}
\end{equation}
where the valve switch takes place at $t = 0$, $H\left(t\right)$ is the 
Heaviside unit step function and $C_{1}$ and $C_{2}$ refer to the amplitude 
and base line level of the isotopic step. 
The effect of smoothing can be seen as the 
convolution of $S\left(t\right)$ with a smoothing function $\mathcal{G}$.
\begin{equation}
\label{Eq.7}
m\left(t\right) = \left[S \ast \mathcal{G} \right] \left(t \right)
\end{equation}
where $m \left(t \right)$ is the measured signal.
The derivative of the signal $\frac{dm}{dt}$ 
yields the impulse response of the
system as: 
\begin{equation}
\label{Eq.8}
\frac{dm}{dt} = \frac{dS}{dt} \ast \mathcal{G}=
C_{1}\frac{dH}{dt} \ast \mathcal{G}=
C_{1}\delta_{Dirac} \ast \mathcal{G}
\end{equation}

We fit the obtained data with a scaled version of the cumulative 
distribution function of a 
Log-Normal distribution described as
\begin{equation}
\delta_{model} \left( t \right) = \frac{K_1}{2} \left[ 1 + \mathrm{erf} 
\left( \frac{\ln t - t_{valve}}{S  \sqrt{2}} \right) \right] + K_2
\end{equation}
where we estimate values for $K_1, K_2, t_0 \textrm{ and } S$ by means 
of a least square optimization (figure 7). 
The fit parameters are used to normalise the isotopic step. 
Based on the
latter we then calculate the impulse response of the system  as described in Eq.
\ref{Eq.8}
As $t=t_{valve}=0$ we consider the time at which the normalised step
is equal to 0.5 and accordingly normalise the time scale (figure 8).
The impact of the memory effects on the impulse response is visualised as 
the ratio $\mathcal{R}\left( t \right)/
\mathcal{R} \left( t = 0 \right)$,  calculated for $t\ge0$ (figure 9). 
One can see that  40 sec after 
the introduction of the $\delta _{Dirac}$ pulse its effect on the measured
signal is below 10\%.
Beyond that point the noise level of the measurement does not allow  for any conclusions
regarding the memory effects.

\subsection{Runs with 4 standards - VSMOW Calibration}

Reporting of water isotopic measurements requires a proper calibration of the 
results on the VSMOW scale. This, in combination with the observed instrumental 
drift, implies the necessity for a frequent VSMOW calibration. In the following 
experiment we inject 5 samples of different isotopic composition 
spanning a range from -9$\permil$ to -54$\permil$ for \delOx 
(-60$\permil$ to -428$\permil$ for \delD). The set of samples consists of 
Copenhagen de-ionised water (CPH-DI)
and a selection of 4 local standards. The latters' isotopic composition has 
been precisely measured with respect to VSMOW and SLAP waters on an IRMS 
 with a high temperature conversion system (HTC).
The water delivery is tuned to 19000 ppm
well within the linear response area. During the whole run ($\approx$ 2 h), 
humidity levels varied with a $\sigma_{\left[ \textrm{H}_{2}\textrm{O} \right]} = 
617 \textrm{ ppm}$. 
 
The raw data of the experiment are presented in figure 10. 
Before any further processing, a humidity calibration of the data
is performed by scaling all data points to the level of 20000 ppm in the same
fashion as described in the previous section. We choose sections of 35 data
points ($\approx 4 \textrm{ min}$) for every separate injection of a sample. 
Based on two of those sections
and in combination with the values obtained by the HTC system we calculate the 
coefficients of a VSMOW calibration line as described by:

\begin{equation}
\label{Eq.5}
\delta_{\textrm{VSMOW}} = 
a_{\textrm{VSMOW}}\cdot\delta_{\textrm{measured}} 
+ b_{\textrm{VSMOW}}
\end{equation}

The sections are carefully selected in order to exclude data affected 
by memory effects occuring for some seconds after the valve switch between samples.
The results of the measurements are presented in table \ref{Table2}. The final 
values are compared to the values of the samples as measured on the HTC mass
spectrometer system. The overall precision of the system for \delD and \delOx
is below 0.5 $\permil$ and 0.1 $\permil$ respectively. The average of the differences between 
the CFA - CRDS and the HTC system is -0.05 $\permil$ for \delOx 
and -0.42 $\permil$ for \delD.

%
%
\section{Conclusions and Outlook}
We have demonstrated the feasibility of on-line 
liquid water isotopic
measurements by interfacing a low volume continuous stream flash evaporator
to a Cavity Ring Down Infra Red spectrometer. 
We have assesed the performance of the system regarding
precision, accuracy, possible instrumental drifts, memory effects 
and dependancy on varrying
humidity levels. 
The observed instrumental
drifts are minimal, thus allowing for reasonable sampling times.
Additionally, the humidity dependance of the system is easily corrected via
a careful and repeatable calibration procedure.

We have also indicated a procedure
to calibrate the measured isotopic ratios on the VSMOW scale using local standard waters.
The system's precision is 
comparable to that of modern mass spectrometry measurement systems tailored
for water isotope analysis.

The use of the system is oriented towards the area of high resolution on-line
continuous isotopic analysis of ice cores. 
The low power consumption and portability,
offer the possibility for field operation. 
The proposed calibration technique can be performed in $\approx$ 30 min, it
requires a small amount of water ($\approx$ 2-3 ml per standard) and can 
in principle be automated.
Dispersion and memory effects are expected to smooth the acquired signals
thus reducing the resolution that can be obtained with this technique. 
Consequently a careful determination of the systems resolution is essential as
a next step towards continuous ice core measurements. Further reduction of 
the volume of the transfer lines prior to the optical cavity can potentially
improve the system to that end.

%
%

\section*{Acknowledgements}
We would like to thank Professor Dorthe Dahl-Jensen for her support in our work.
Jim White and Bruce Vaughn have contributed to this work with valuable 
comments and discussions.
We would also like to thank Aaron van Pelt for his assistance during this project.
This work was partly funded by the Marie Curie RTN "Network for Ice Sheet 
and Climate Evolution",  MRTN-CT-2006-036127


\bibliographystyle{IEEEtran}


\newpage

\begin{table*}
\begin{center}
\begin{tabular*}{0.9\textwidth}{@{\extracolsep{\fill}}l r r r } \toprule
\multicolumn{1}{l}{\hspace{2mm}Sample} & 
\multicolumn{1}{c}{\delD} &
\multicolumn{1}{c}{$\bar{\delta}\textrm{D}$} & 
\multicolumn{1}{c}{$\delta\textrm{D}_{\textrm{HTC}}$}\\
\multicolumn{1}{l}{\hspace{2mm}ID}     &
\multicolumn{1}{c}{[$\permil$]} & 
\multicolumn{1}{c}{Span} &
\multicolumn{1}{c}{$\delta\textrm{D}_{\textrm{HTC}} - 
\bar{\delta} \textrm{D}$}\\
\midrule
\hspace{4mm}CPH-DI & -60.33 & -60.26 & -\hspace{4mm}\\
\hspace{4mm}CPH-DI & -60.18 & 0.1 & -\hspace{4mm}\\
\hspace{4mm}-22 & -167.66 & -168.03 & -168.4\hspace{4mm}\\
\hspace{4mm}-22 & -168.40 & 0.52 & -0.37\hspace{4mm}\\
\hspace{4mm}Crete & -260.70 & -260.94 & -261.9\hspace{4mm}\\
\hspace{4mm}Crete & -261.19 & 0.34 & -0.96\hspace{4mm}\\
\hspace{4mm}-40 & -309.13 & -309.57 & -310\hspace{4mm}\\
\hspace{4mm}-40 & -310.00 & 0.62 & -0.44\hspace{4mm}\\
\hspace{4mm}DC02 & -427.31 & -427.58 & -427.5\hspace{4mm}\\
\hspace{4mm}DC02 & -427.85 & 0.38 & 0.08\hspace{4mm}\\
\midrule
\hspace{4mm}Std. dev all &  &  & \textbf{0.39}\\
\hspace{4mm}$\delta\textrm{D}_{\textrm{HTC}} - 
\bar{\delta} \textrm{D all} $ & & & \textbf{-0.42}
\end{tabular*}
\begin{tabular*}{0.9\textwidth}{@{\extracolsep{\fill}}l r r r } \toprule
\multicolumn{1}{l}{\hspace{2mm}Sample} & 
\multicolumn{1}{c}{\delOx} & 
\multicolumn{1}{c}{$\bar{\delta}^{18}\textrm{O}$} &
\multicolumn{1}{c}{$\delta^{18}\textrm{O}_{\textrm{HTC}}$}\\
\multicolumn{1}{l}{\hspace{2mm}ID}     &
\multicolumn{1}{c}{[$\permil$]} &
\multicolumn{1}{c}{Span} & 
\multicolumn{1}{c}{$\delta^{18}\textrm{O}_{\textrm{HTC}} - 
\bar{\delta} ^{18}\textrm{O}$}\\
\midrule
\hspace{4mm}CPH-DI & -8.69 & -8.70 & -\hspace{4mm}\\
\hspace{4mm}CPH-DI & -8.70 & 0.01 & -\hspace{4mm}\\
\hspace{4mm}-22 & -21.9 & -21.92 & -21.9\hspace{4mm}\\
\hspace{4mm}-22 & -21.93 & 0.02 & 0.02\hspace{4mm}\\
\hspace{4mm}Crete & -33.57 & -33.54 & -33.64\hspace{4mm}\\
\hspace{4mm}Crete & -33.52 & 0.04 & -0.1\hspace{4mm}\\
\hspace{4mm}-40 & -39.69 & -39.83 & -39.97\hspace{4mm}\\
\hspace{4mm}-40 & -39.98 & 0.2 & -0.14\hspace{4mm}\\
\hspace{4mm}DC02 & -54.08 & -54.11 & -54.08\hspace{4mm}\\
\hspace{4mm}DC02 & -54.14 & 0.04 & 0.03\hspace{4mm}\\
\midrule
\hspace{4mm}Std. dev all &  &  & \textbf{0.06}\\
\hspace{4mm}$\delta^{18}\textrm{O}_{\textrm{HTC}} - 
\bar{\delta} ^{18}\textrm{O all} $ & & & \textbf{-0.05}\\
\bottomrule
\end{tabular*}
\end{center}
\caption{Results of the 4 Standards Experiment}

\label{Table2}
\end{table*}
\newpage

\begin{figure*}
\centering
\includegraphics[width = 130mm]{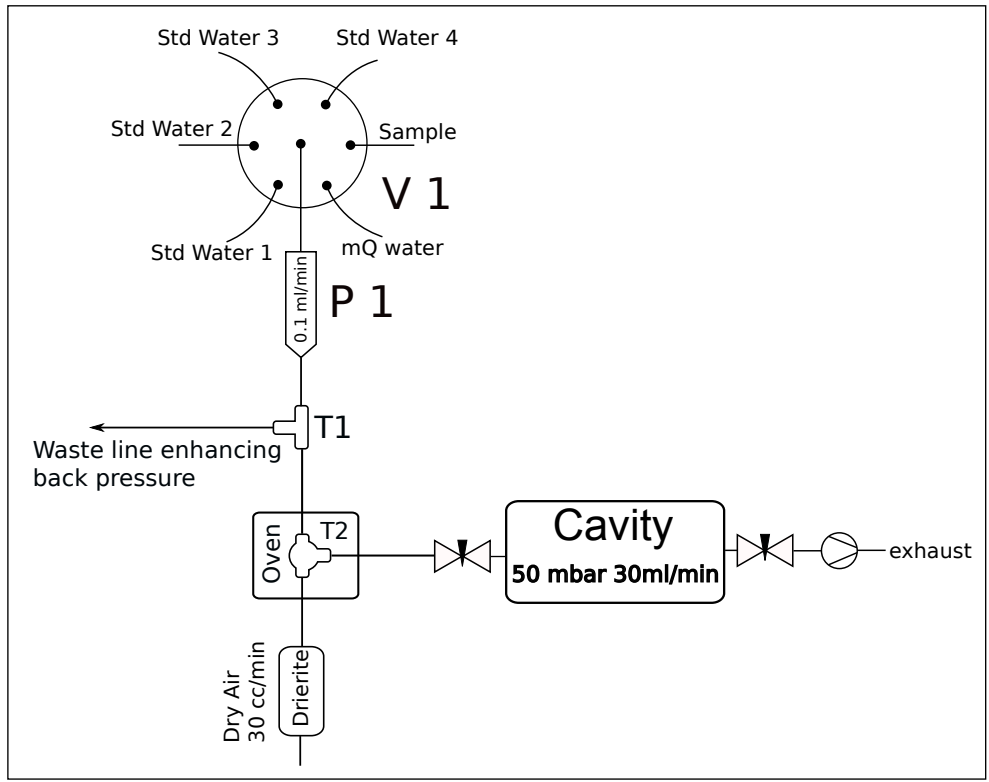}
\caption{Block diagram of the CFA-CRDS system}
\label{Fig1}
\end{figure*}

\pagebreak

\begin{figure*}
\centering
\includegraphics[width = 130mm]{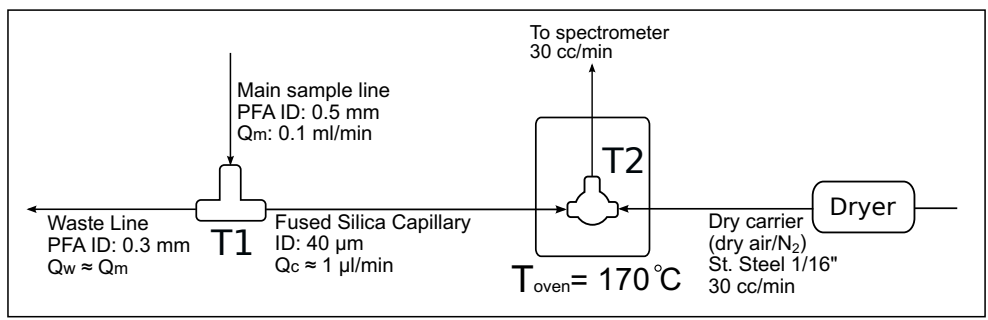}
\caption{Detailed section of the sample split}
\label{block1a}
\end{figure*}

\pagebreak

\begin{figure*}
\centering
\includegraphics[width = 130mm]{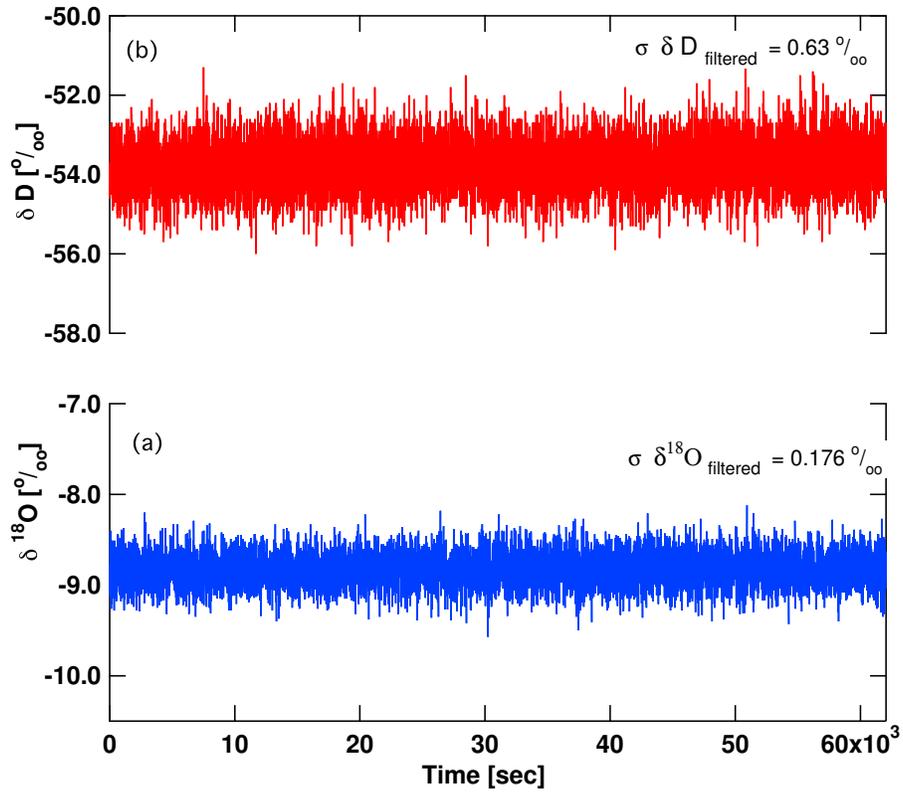}
\caption{Injection of de-ionized water over the period of 17 hours. Results for 
\delOx (a) and \delD (b)}
\label{Fig2}
\end{figure*}
\pagebreak

\begin{figure*}
\centering
\includegraphics[width = 130mm]{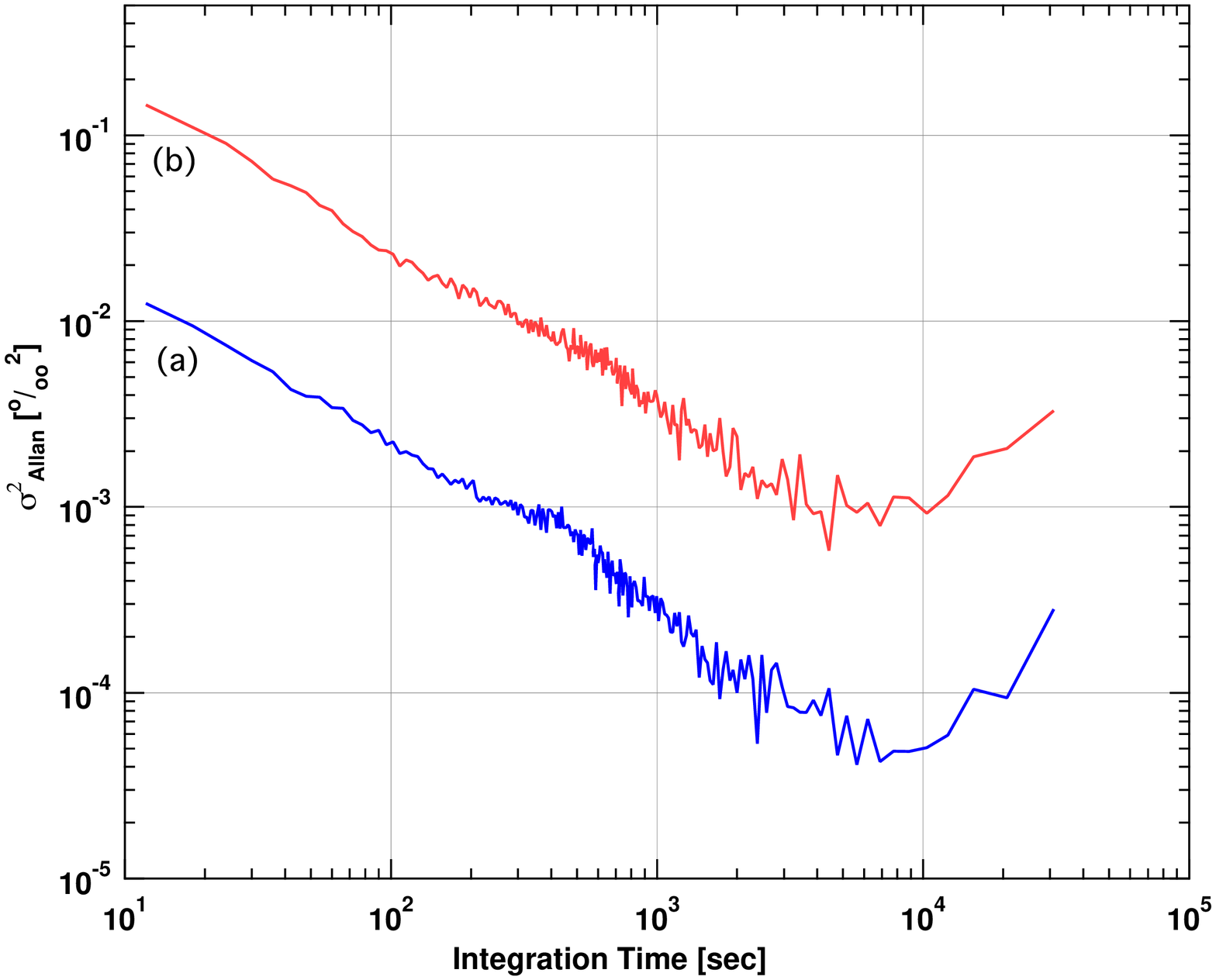}
\caption{Allan variance plots for \delOx (a) and \delD (b) from the data in 
figure \ref{Fig2}.}
\label{Fig3}
\end{figure*}

\pagebreak

\begin{figure*}
\begin{center}
\includegraphics[width = 130mm]{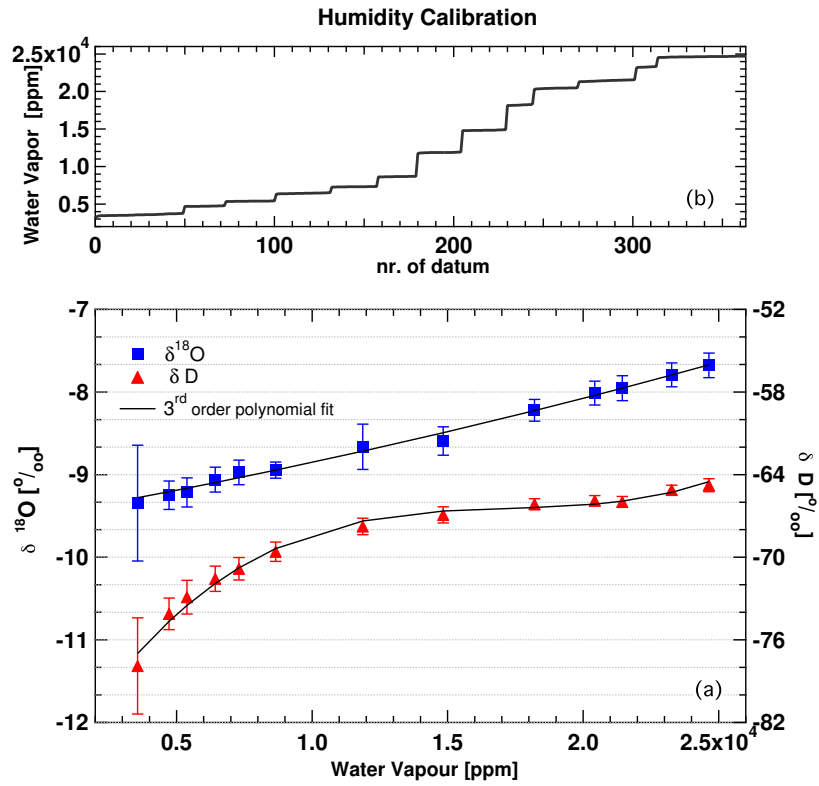}
\caption{Humidity calibration. The averages of the isotopic values for each section
are plotted as blue squares for \delOx and red triangles for \delD in graph
(a). Data are fitted with a 3rd order polynomial regression model (solid
lines).
The error bars represent $\pm 1 \sigma$ of each processed section.
In (b) we plot the humidity levels of each section.}
\label{Fig4}
\end{center}
\end{figure*}
\pagebreak

\begin{figure*}
\begin{center}
\includegraphics[width = 130mm]{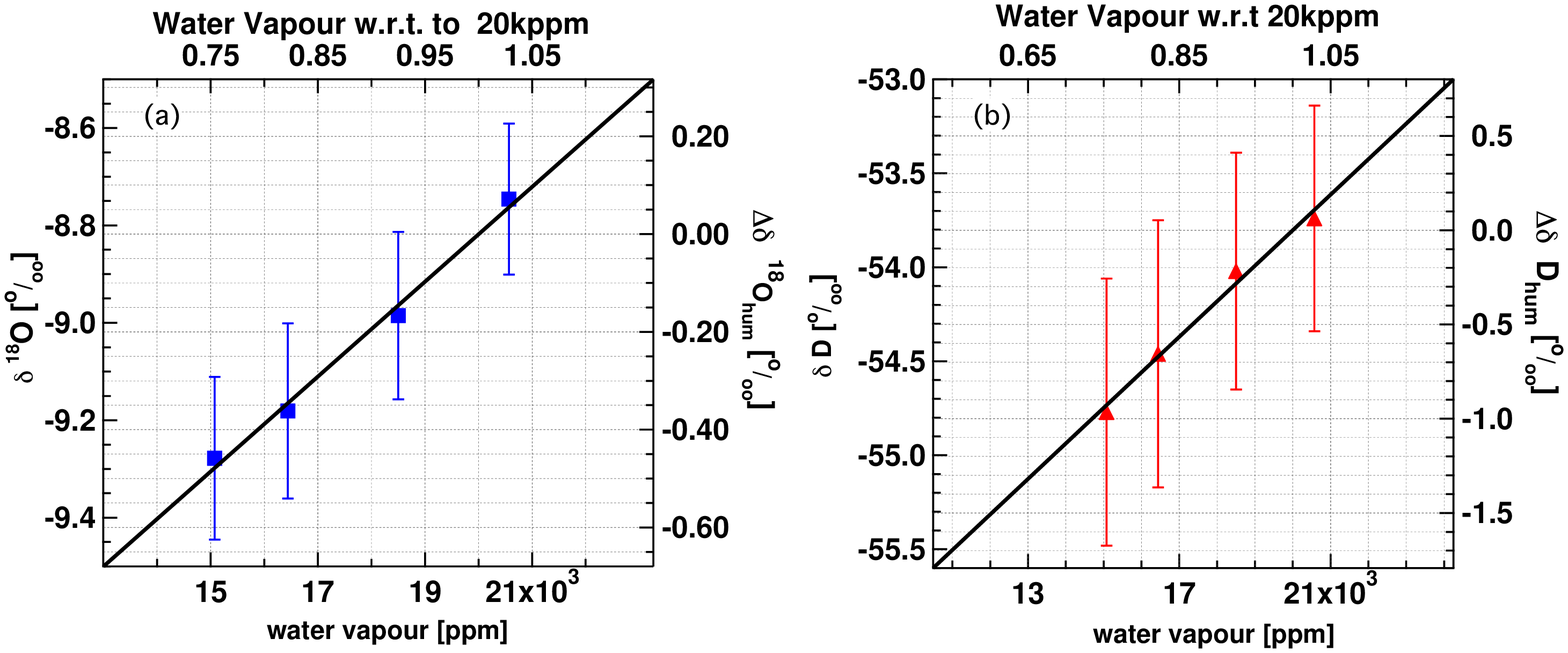}
\caption{The averages of each humidity 
level section are plotted
as blue squares and red triangles for \delOx and \delD respectively. 
Error bars represent a
$\pm 1 \sigma$ of each section. }
\label{Fig5}
\end{center}
\end{figure*}
\pagebreak

\pagebreak

\begin{figure*} 
\begin{center} 
\begin{minipage}{140mm} 
\subfigure[]{ 
\resizebox*{7cm}{!}{\includegraphics{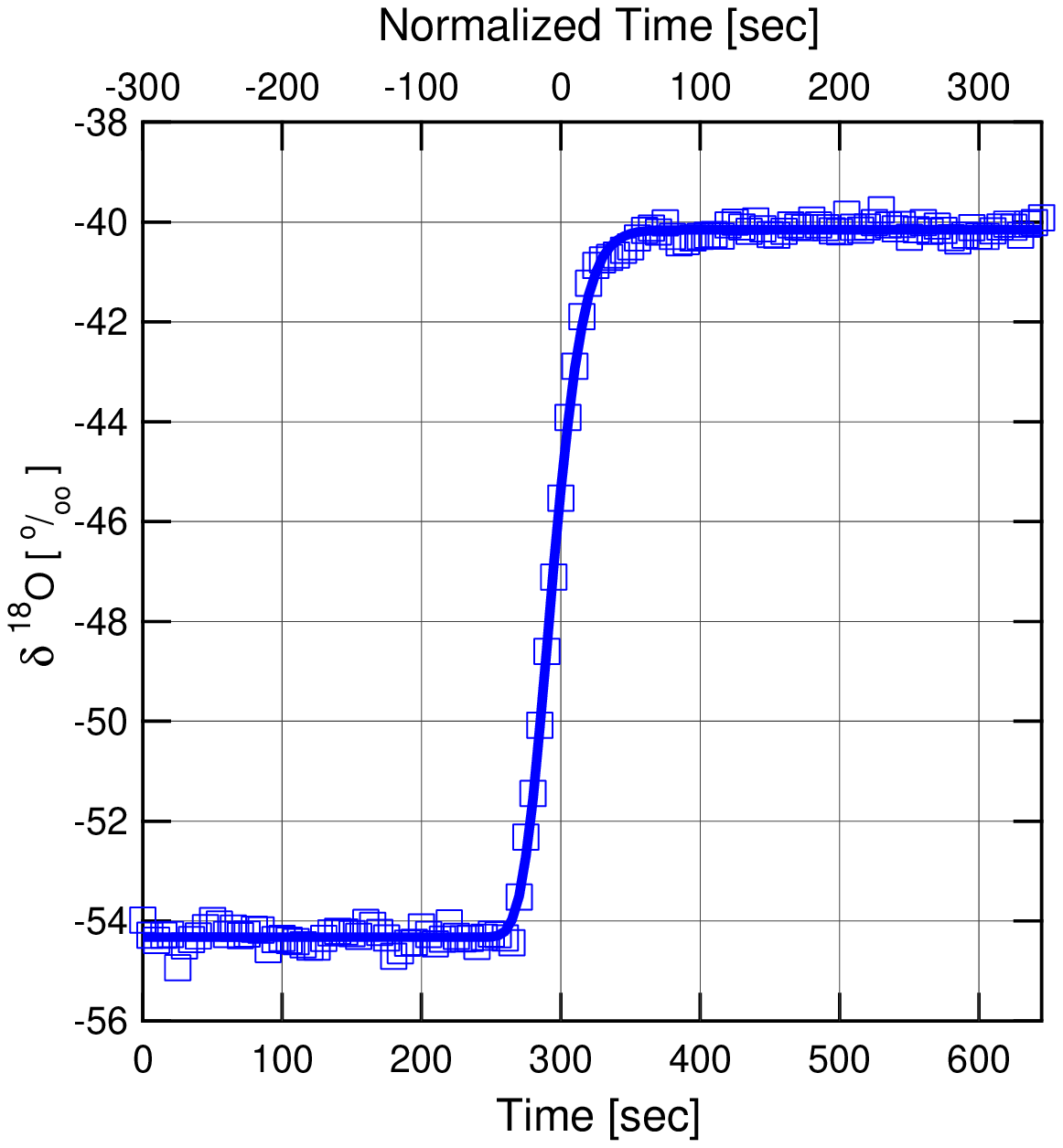}}}%
\subfigure[]{ 
\resizebox*{7cm}{!}{\includegraphics{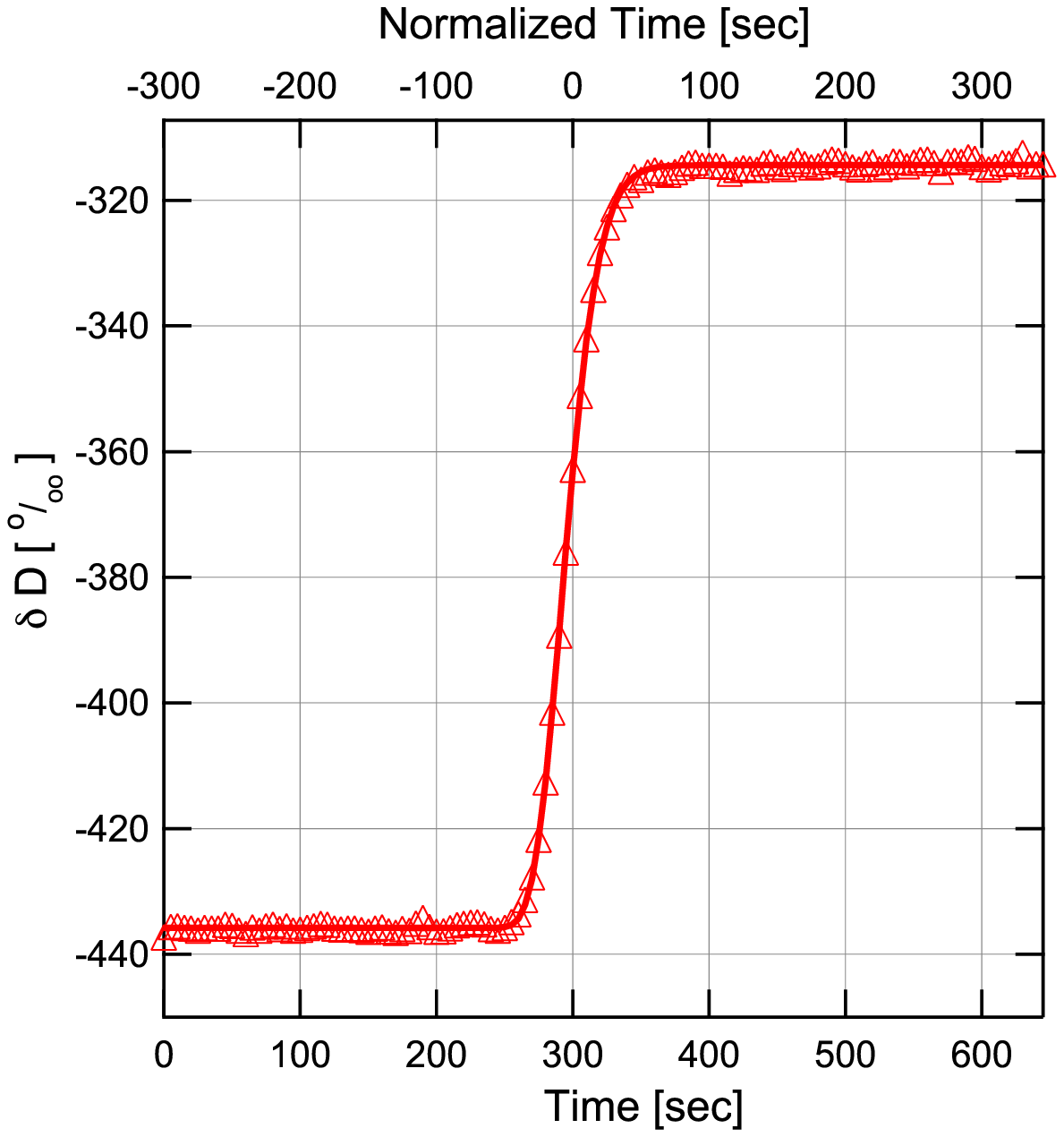}}}%
\caption{ Raw \delOx and \delD data during the valve switch 
between standard water
``DC02'' and ``-40'' are plotted with blue squares and red triangles
respectively.
We fit a Log Normal distribution model plotted with blue (a)for \delOx
and red (b) for \delD..}%
\label{cdfs} 
\end{minipage} 
\end{center} 
\end{figure*}

\pagebreak
\begin{figure*} 
\begin{center} 
\begin{minipage}{140mm} 
\subfigure[]{ 
\resizebox*{7cm}{!}{\includegraphics{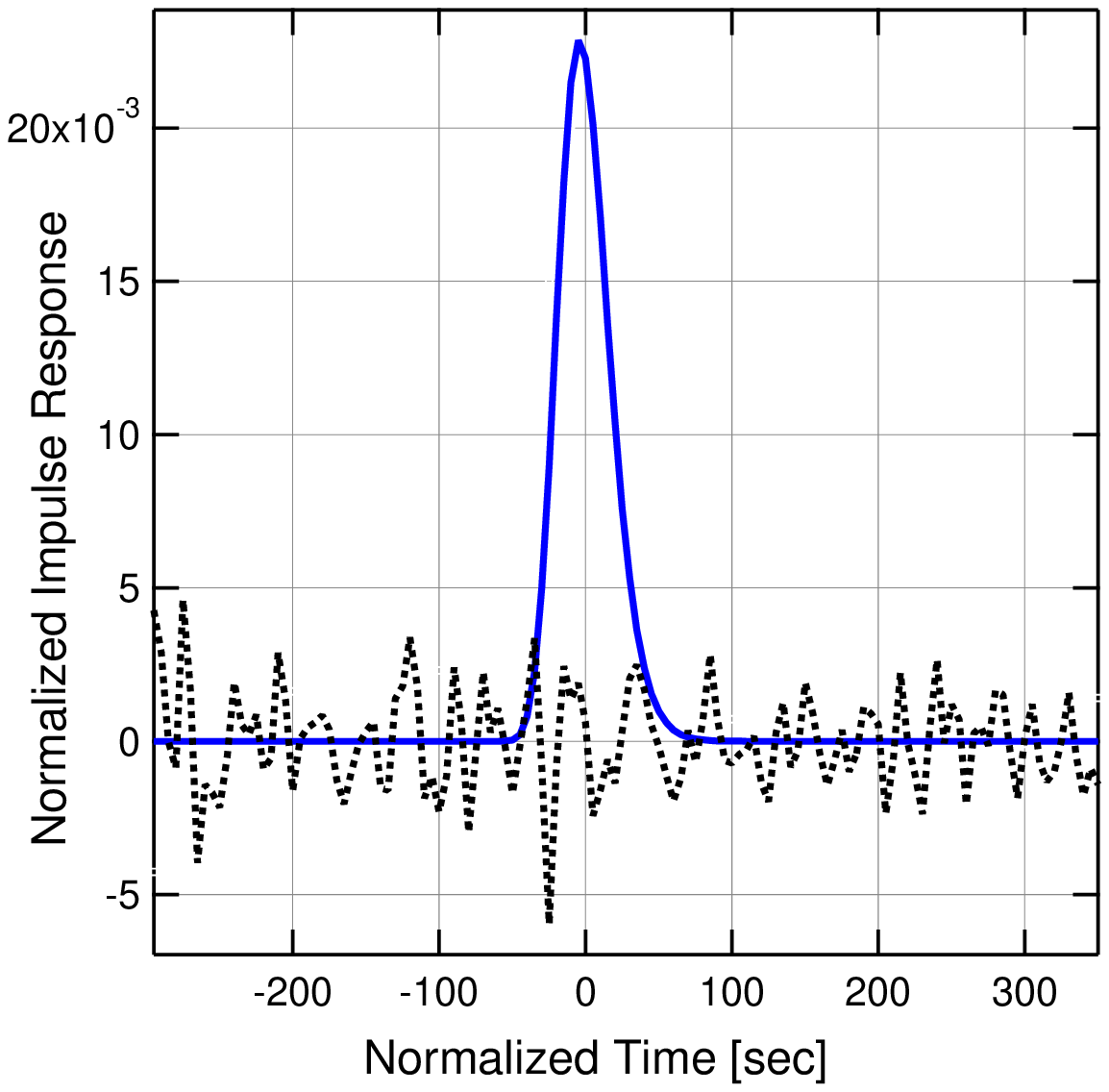}}}%
\subfigure[]{ 
\resizebox*{7cm}{!}{\includegraphics{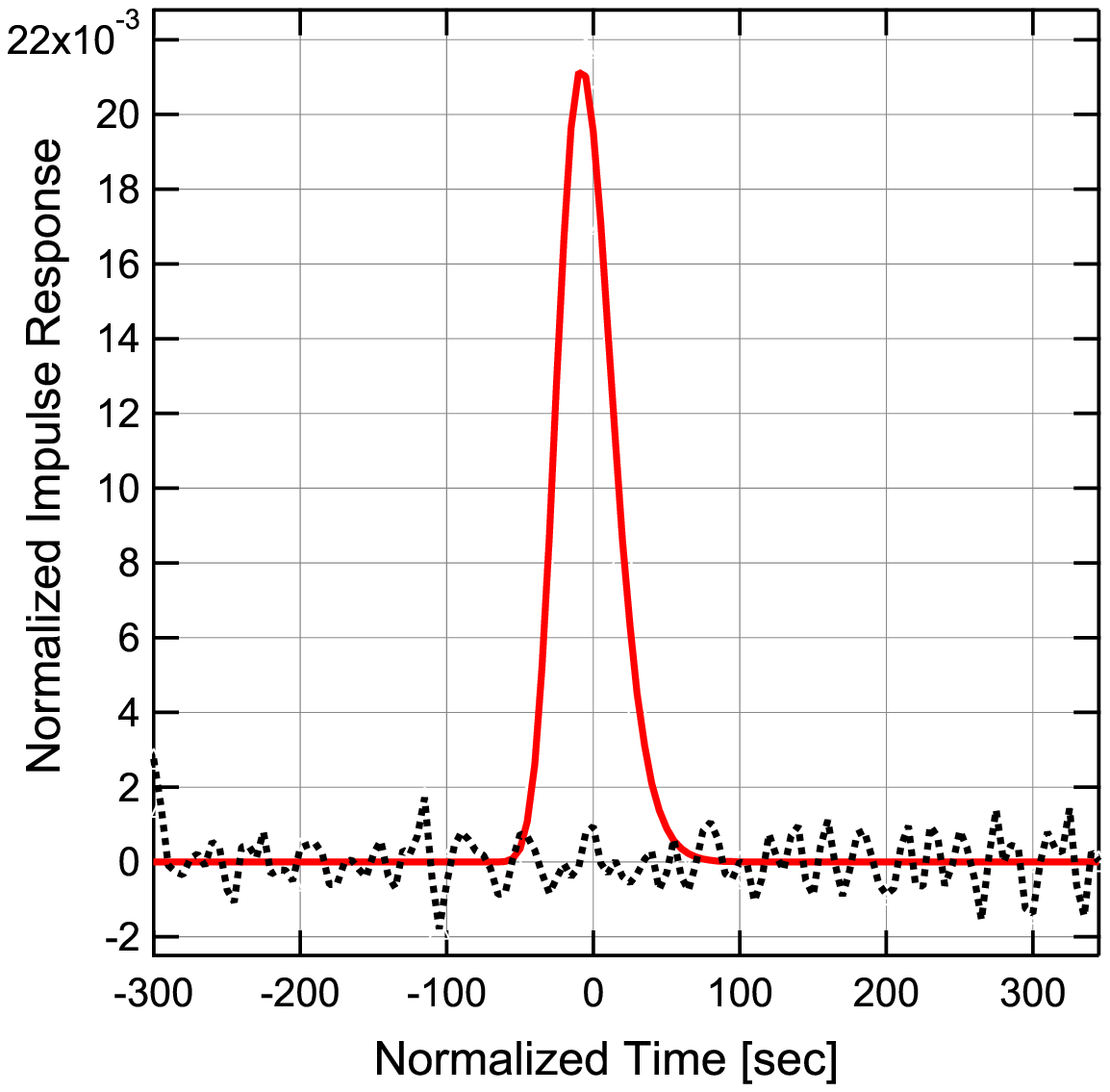}}}%
\caption{Response of the system on an isotopic $\delta_{Dirac}$ pulse for
\delOx (blue curve) and \delD (red curve) introduced at $t=0$.
The dotted lines represend the residuals of the fit.}%
\label{response_dirac} 
\end{minipage} 
\end{center} 
\end{figure*} 

\pagebreak
\begin{figure*} 
\begin{center} 
\begin{minipage}{140mm} 
\subfigure[]{ 
\resizebox*{7cm}{!}{\includegraphics{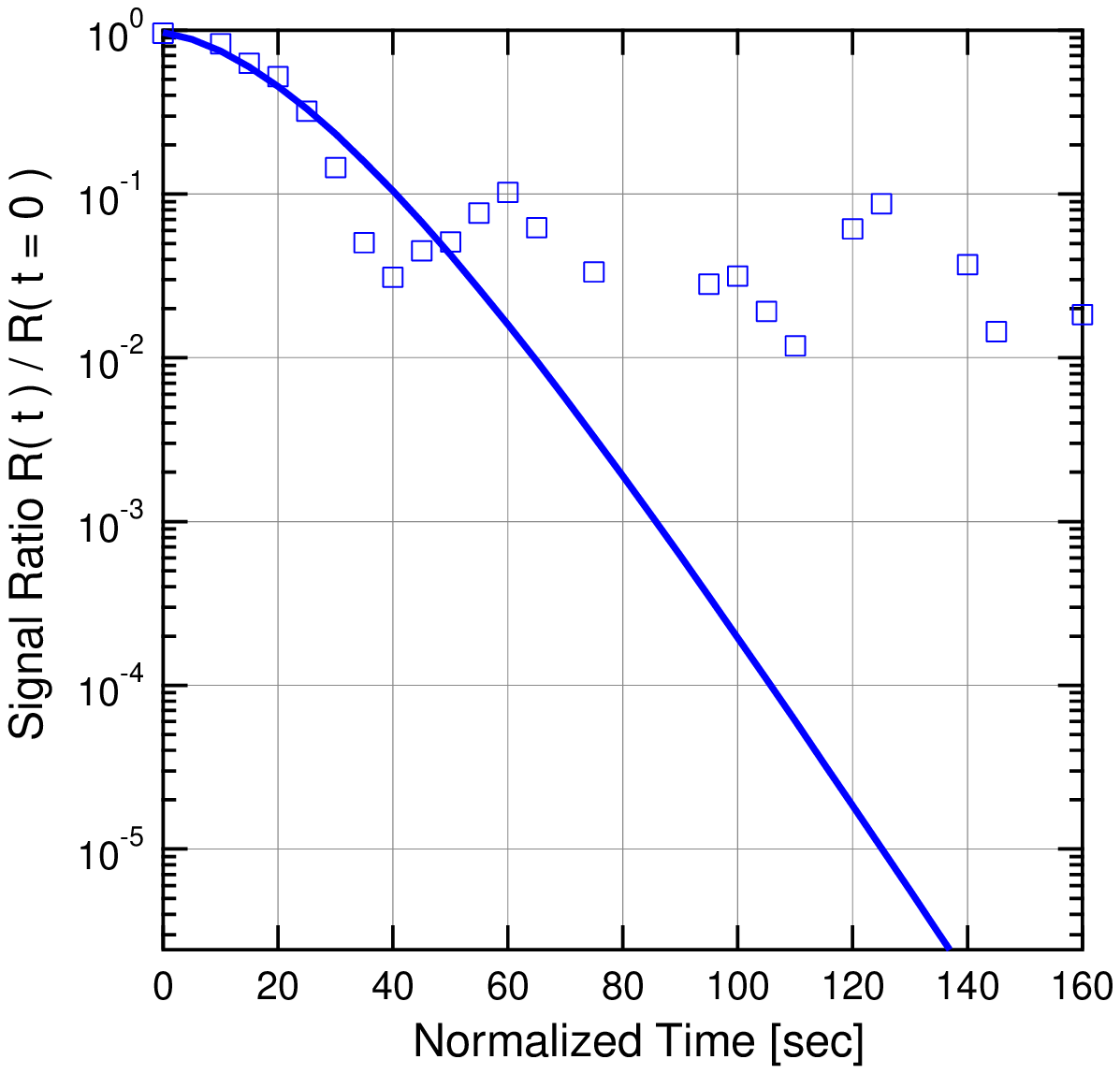}}}%
\subfigure[]{ 
\resizebox*{7cm}{!}{\includegraphics{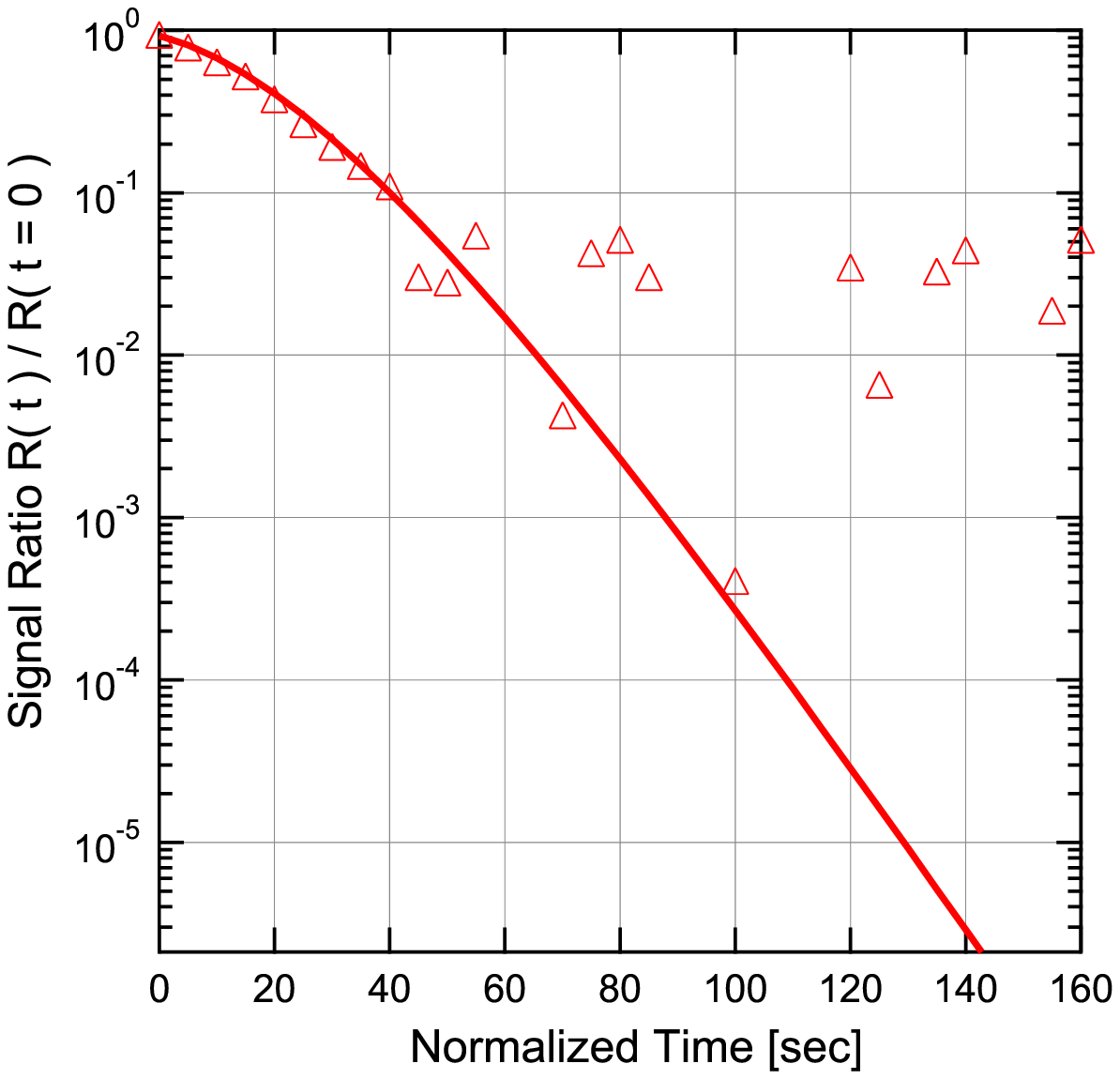}}}%
\caption{Quantification of the memory effect for \delOx (a)
and \delD (b).
$R \left( t \right)$ refers to the amplitude of the impulse response 
of the system at time $t$. }%
\label{memory} 
\end{minipage} 
\end{center} 
\end{figure*} 

\end{document}